# Na$_9$Bi$_5$Os$_3$O$_{24}$: A Unique Diamagnetic Oxide Featuring a Pronouncedly Jahn-Teller Compressed Octahedral Coordination of Osmium(VI)

Gohil S. Thakur,[†,‡] Hans Reuter,[⸸] Alexey V. Ushakov,[‖] Gianpiero Gallo,[⊓] Jürgen Nuss,[⊓] Robert E. Dinnebier,[⊓] Sergey V. Streltsov,[‖] Daniel I. Khomskii,[§] and Martin Jansen[⊓,*]

[†] Max Planck Institute for Chemical Physics of Solids, Nöthnitzerstr. 40, 01187 Dresden, Germany

[‡] Faculty of Chemistry and Food Chemistry, Technical University, 01069 Dresden, Germany

[⸸] Institute for Chemistry of new Materials, University of Osnabrück, Barbarastraße 7, 49069 Osnabrück, Germany

[‖] M. N. Mikheev Institute of Metal Physics, Ural Branch of Russian Academy of Sciences, 620002 Ekaterinburg, Russia

[⊓] Max Planck Institute for Solid State Research, Heisenbergstr. 1, 70569 Stuttgart, Germany

[§] II Physikalisches Institut, Universitaet zu Köln, Zülpicher Str. 77, 50937 Köln, Germany

KEYWORDS. Jahn-Teller compression, spin-orbit coupling, multinary osmate, hydrothermal synthesis.

**ABSTRACT:** The Jahn-Teller (JT) theorem constitutes one of the most popular and stringent concepts, applicable to all fields of chemistry. In open shell transition elements chemistry and physics, $3d^4$, $3d^9$, and $3d^7$(low-spin) configurations in octahedral complexes serve as particular illustrative and firm examples, where a striking change ("distortion") in local geometry is associated to a substantial reduction of electronic energy. However, there has been a lasting debate, almost of a historical dimension, about the fact that the octahedra are found to exclusively elongate, (at least for $e_g$ electrons). Against this background, the title compound displays two marked features, (1) the octahedron of oxygen atoms around Os$^{6+}$ with $d^2$ configuration is drastically compressed, in contrast to the standard JT expectations, and (2) the splitting of the $t_{2g}$ set of $5d$ orbitals induced by this compression is extreme, such that a diamagnetic ground state results. What we see is obviously a Jahn-Teller distortion resulting in a *compression* of the respective coordinating octahedron and acting on the $t_{2g}$ set of orbitals. Both these issues are unprecedented. Noteworthy, the splitting into a lower $d_{xy}$ (hosting two $d$ electrons with opposite spin) and two higher $d_{xz}$ and $d_{yz}$ orbitals is so large that for the first time ever the Hund´s coupling for $t_{2g}$ electrons is overcome. We show that these effects are not forced by structural frustration, the structure offers sufficient space for Os to shift the apical oxygen atoms to a standard distance. Local electronic effects appear to be responsible, instead. The relevance of these findings is far reaching, since they provide insights in the hierarchy of perturbations defining ground states of open shell electronic systems. The four component system studied here, offers substantially more structural and compositional degrees of freedom, such that a configuration could form that enables Os$^{6+}$ to adopt its apparently genuine diamagnetic ground state.

## INTRODUCTION

Dealing with multitudes of particles constitutes an inescapable and major challenge in chemistry, even within the framework of the Born-Oppenheimer approximation. While for molecular species the numbers of atoms involved are finite, they are virtually infinite for extended solids. In coping with the task of structuring and classifying chemical knowledge for purposes of e.g. communication, teaching, understanding or planning of chemical research the enormous richness of empirical and computational results need to be cast into comprehensible concepts.[1] Along the most popular of such approaches, one tries to structure observations, and even to derive cause-effect relationships, by breaking down collective properties into additive increments. Prominent examples are atomic radii (from distances),[2] electronegativity (from shortening of covalent bonds due to polarity),[3] oxidation states (assigning charges by systematic definition),[4] or the reverse, bulk diamagnetic response of a compound (from calculated increments).[5] Such procedures are most reliable for compounds consisting of only two different elements, making it easier to separate the individual contributions of the constituting atoms. However, as a distinct disadvantage of such binaries the degrees of freedom in evolving different compositions, structures, bonding schemes and properties are substantially limited. With respect to both these issues, the opposite holds true for multinary compositions, and the individual constituting atom types are less constrained in striving for their respective pertinent preferred configurations.

Here we report on an illustrative example of a novel four-component oxide Na$_9$Bi$_5$Os$_3$O$_{24}$ in which the additional degrees of freedom have enabled a complex composition and a singular crystal structure where the constituents Na, Bi, Os, and O come close to their characteristic structure and bonding properties. As a surprising result, although containing Os$^{6+}$ with a $5d^2$ electron configuration, the compound is diamagnetic and exhibits a drastically compressed octahedron around osmium, which causes an exceptionally strong crystal field splitting of the $t_{2g}$ levels thus lifting the degeneracy of unequally occupied orbitals according to a Jahn-Teller (JT) effect. This finding is very special in another, more general respect. In known solids the transition metal (TM) ions with partially-filled $t_{2g}$ levels always obey the first Hund's rule, *i.e.* electrons are arranged such that the maximum S results. We are not aware of any exception of this rule for $t_{2g}$ shells in extended solids. In one incident, La$_4$Ru$_2$O$_{10}$, it was initially suggested that Ru$^{4+}$ has an S = 0 ground state.[6] This was challenged later, and shown that the nonmagnetic character of this system is due to the formation of singlet Ru dimers ("molecules in solids").[7] In our case, as we show below, Os$^{6+}$($5d^2$) really has a singlet ground state with S = 0, strongly violating the first Hund's rule. This particular configuration may form because of a synergetic action of weaker structural matrix effects of the many-component material and a strong local JT effect, seen for the first time to act on the $t_{2g}$ levels of an octahedral complex and to induce a compression.

## EXPERIMENTAL SECTION

**Synthesis and crystal growth.** Intimately mixed powders of Bi$_2$O$_3$ (0.0777 g, 0.5 mol, Alfa Aesar 99.9 %) OsO$_2$ (0.1482, 2 mol, Alfa Aesar 99.5%) and Na$_2$O$_2$ (0.078 g, 3 mol, homemade) to which 0.5 ml of water was added (1 ml of 5 M NaOH solution can be used instead of $Na_2O_2$), were heated under high oxygen pressure and at high temperature. The solid starting materials were ground in an Ar filled dry glove box and transferred into a gold finger welded at one end. Water was added dropwise outside the glove box immediately before crimping the tube from the top end and placing it in a steel autoclave. The autoclave was tightly sealed after approximately 11.2 ml of liquid oxygen was condensed in it. This amount of oxygen would generate a pressure of around 350 MPa at 773 K. The reaction was carried out for 4 days in a vertical furnace before cooling it naturally. The product was vacuum filtered inside a fume hood and rinsed with distilled water first and subsequently with ethanol. Black reflective hexagonal blocks or thick plates were obtained which were further cleaned and separated by ultra-sonication under ethanol. Crystals of maximum dimensions of up to $1 \times 1 \times 1$ mm$^3$ could be obtained. A white powdery layer on the crystals appears after a few weeks of exposure to moist air, while the powder diffraction pattern did not seem to change. The product was stored in a glove box.

Caution: Under the given conditions, reaction involving osmium generates highly toxic $OsO_4$, hence, care must be taken during the release of pressure inside the fume hood. Before fetching the product, the opened autoclave was left for several hours inside the fume hood to allow $OsO_4$ to escape completely. Use of proper PPE (eye protection, respirator as well as latex gloves) is necessary.

**Chemical analyses.** Semiquantitative analysis was performed by scanning electron microscopy-energy dispersive analysis by X-rays (SEM-EDAX). Many crystals from different batches were examined and the data averaged to obtain the metal composition, see Figure S1 in supporting information (SI). Wet chemical analyses of the metal contents were performed by Mikroanalytisches Labor Pascher, An der Pulvermühle 1, D-53424 Remagen, Germany (microwave pressure digestion with $HNO_3$/HF/HCl; Bi and Os determined by ICP-OES, Na by AAS). Two different runs on the same batch of sample were averaged to obtain the atomic fraction. The results are presented Table S1 in SI.

Thermogravimetric (TGA) analysis of the collected single crystals was carried out on a Netzsch STA 449 C analyzer. About 30 mg of the sample was placed in a corundum crucible, which was heated and subsequently cooled at a rate of 5 K min$^{-1}$ in the range of 300–1273 K under dynamic argon flow, see Figure S2 in SI.

**Physical properties.** Susceptibility of loose single crystals was measured in applied magnetic fields $\mu_0 H$ = 0.1, 1.0 and 3.5 T and in the temperature range between 2 and 350 K in a MPMS-XL7 magnetometer (Quantum Design). Temperature dependent resistivity in the temperature range, $T$ = 150−400 K was measured on a block-like single crystal of dimension approximate dimensions $1 \times 0.8 \times 0.7$ mm$^3$ using a two-probe method in a PPMS instrument (Quantum Design).

**Crystal structure determination.** The determination of the crystal structure by X-ray diffraction did not proceed straight forwardly and required to employ both powder and single crystal techniques in an alternating fashion.

X-Ray Powder Diffraction (XRPD) measurements were performed using a Stoe Transmission Powder Diffraction System (STADI-P, STOE & CIE, Ge(111) Johansson-type monochromator, AgKα$_1$ radiation (λ = 0.55941 Å)) that was equipped with an array of three linear position-sensitive MYTHEN 1K detectors from Dectris Ltd. of approximately 18° 2θ opening angle each. The finely powdered sample of $Na_9Bi_5Os_3O_{24}$ was placed in a glass capillary of 0.3 mm (Hilgenberg glass No. 14) and spun during measurement for improving particle statistics. The measurement in the range from 1.0 – 111.0° 2θ with a step width of 0.015° 2θ took 3 hrs (Figure 1). For indexing of the powder pattern of $Na_9Bi_5Os_3O_{24}$ at $T$ = 298 K, the program TOPAS version 6 (Bruker-AXS, 2018)[8] was used, leading to a hexagonal unit cell with parameters of $a$ = 9.8264(1) and $c$ = 12.8573(2) Å ($V$ = 1075.16(3) Å$^3$). The most probable space groups were determined as $P31c$ (159), $P\bar{3}1c$ (163), $P6_3mc$ (186), $P\bar{6}2c$ (190), and $P6_3/mc$ (194) from the observed extinction rules, out of these $P\bar{6}2c$ was confirmed after structure determination. Structure determination of $Na_9Bi_5Os_3O_{24}$ was performed in all possible space groups by the method of Charge Flipping[9], supported by the inclusion of the tangent formula[10] as implemented in TOPAS.[8] The positions of the heavier atoms (Bi, Os) and some candidate for the sodium atoms were found for several space groups but with a clear preference for $P\bar{6}2c$. The space group assumed and the heavy atom structure were confirmed using single crystal X-ray diffraction data, which also enabled to identify the missing light atoms. These final results comply well with the PXRD, as was validated by Rietveld refinement using the TOPAS program. An overall isotropic temperature factor was refined. The profiles related to the final Rietveld refinement are shown in Figure 1. The weighted profile R-factor is 3.55 %, the Bragg R-factor is 1.91 %, with a goodness of fit of 2.26. The atomic coordinates are given in Table S2 and a selection of intramolecular distances and angles is given in Table S3 in SI.

Crystals suitable for single-crystal X-ray diffraction were selected under highly viscous oil, and mounted with grease on a loop made of Kapton foil (Micromounts™, MiTeGen, and Ithaca, NY). Diffraction data were collected at 298 K with a SMART APEXII CCD X-ray diffractometer (Bruker AXS, Karlsruhe, Germany), using graphite-monochromated Mo-Kα radiation. Reflection intensities were integrated with the SAINT subprogram in the Bruker Suite software,[11] a multi-scan absorption correction was applied using SADABS,[12] and the structure was refined by full-matrix least-square fitting with the SHELXTL software package.[13,14] According to the systematic reflection condition $hh\bar{2h}l$ only present for $l$ = 2n, trigonal space groups $P31c$ (159), $P\bar{3}1c$ (163), and hexagonal space groups $P6_3mc$ (186), $P\bar{6}2c$ (190) and $P6_3/mmc$ (194) have been checked for structure solution, however, any structure solution method of the SHELXS package failed. The heavy-element positions from powder solutions in the non-centrosymmetric space group $P\bar{6}2c$ proved to be a suitable starting model for resolving the complete structure via Difference Fourier analyses. Refinement with anisotropic thermal parameters for all atoms without applying any constraints converged easily. Crystal data and data collection details, positional and isotropic thermal parameters and bond distances, respectively, are given in Tables 1-3, the anisotropic thermal parameters in Table S4 in SI. The crystallographic data have been deposited at ICSD under CSD-No. 2063496.

**Table 1. Crystal and structure refinement data for $Na_9Bi_5Os_3O_{24}$ from SCXRD.**

| | |
|---|---|
| Empirical formula | $Na_9Bi_5Os_3O_{24}$ |
| Formula weight [g/mol] | 2206.41 |
| T [K] | 296(2) |
| Crystal system, space group | Hexagonal, $P\bar{6}2c$ |
| a [Å] | 9.8115(3) |
| c [Å] | 12.8457(5) |
| V [Å$^3$] | 1070.93(8) |
| Z, $d_{calc}$ [g/cm$^3$] | 2, 6.842 |
| $\mu$(MoK$\alpha$) [mm$^{-1}$] | 58.944 |
| F(000) | 1868 |
| $2\Theta_{max}$ | 70° |
| Reflections collected | 75234 |
| Reflections unique, $R_{int}$ | 1637, 0.0877 |
| Data / restraints / parameters | 1637 / 0 / 70 |
| Goodness-of-fit on F$^2$ | 1.040 |
| $R_1$, wR2 [I > 2s(I)] | 0.0208, 0.0529 |
| $R_1$, wR2 [all data] | 0.0268, 0.0567 |
| Absolute structure parameter | -0.038(6) |
| Extinction coefficient | 0.00102(6) |
| $\pm\Delta$ [eÅ$^{-3}$] | 1.728/-3.396 |

**Table 2: Atomic coordinates ($\times 10^4$) and equivalent isotropic displacement parameters (Å$^2 \times 10^3$) for $Na_9Bi_5Os_3O_{24}$ obtained from single crystal structure refinement. U(eq) is defined as one third of the trace of the orthogonalized $U_{ij}$ tensor.**

| Atom | site | x | y | z | $U_{eq}$ |
|---|---|---|---|---|---|
| Os1 | 6h | 6682(1) | 9975(1) | 2500 | 3.9(1) |
| Bi1 | 6g | 6712(1) | 10000 | 5000 | 4.5(1) |
| Bi2 | 4f | 3333 | 6667 | 6051(1) | 7.4(1) |
| Na1 | 2b | 10000 | 10000 | 2500 | 17(1) |
| Na2 | 12i | 2976(6) | 9912(4) | 3789(2) | 13(1) |
| Na3 | 4f | 3333 | 6667 | 3731(4) | 16(1) |
| O1 | 6h | 8035(11) | 12028(12) | 2500 | 10(1) |
| O2 | 6h | 5648(8) | 7895(11) | 2500 | 10(1) |
| O3 | 12i | 5336(5) | 7486(6) | 5037(3) | 7.3(8) |
| O4 | 12i | 5443(6) | 10127(6) | 3631(4) | 8.1(9) |
| O5 | 12i | 7987(7) | 9862(7) | 3703(4) | 8.3(9) |

**Table 3. Selected bond distances for $Na_9Bi_5Os_3O_{24}$.**

| Atom pairs | Distance (Å) | Atom pairs | Distance (Å) |
|---|---|---|---|
| Os1-O2 (×1) | 1.767(10) | Na1-O5 (×6) | 2.457(5) |
| Os1-O1 (×1) | 1.773(11) | Na2-O4 (×1) | 2.335(7) |
| Os1-O4 (×2) | 1.944(5) | Na2-O1 (×1) | 2.406(9) |
| Os1-O5 (×2) | 2.044(5) | Na2-O3 (×1) | 2.440(6) |
| Bi1-O5 (×2) | 2.129(5) | Na2-O5 (×1) | 2.441(7) |
| Bi1-O3 (×2) | 2.140(5) | Na2-O2 (×1) | 2.480(8) |
| Bi1-O4 (×2) | 2.191(5) | Na2-O3 (×1) | 2.490(6) |
| Bi2-O3 (×3) | 2.150(5) | Na3-O2 (×3) | 2.396(6) |
| Bi2-O1 (×3) | 2.937(5) | Na3-O3 (×3) | 2.525(6) |

**Computational methods.** We used the Vienna Ab-initio Simulation Package (VASP) and the generalized gradient approximation for DFT calculations.[15] The on-site Coulomb repulsion $U$ and Hund's intra-atomic exchange $J_H$ parameters were chosen to be 1.1 and 0.5 eV, so that $U$-$J_H$ = 0.6 eV.[16] A 5×5×5 mesh in $k$-space was used. For simplicity we assumed ferromagnetic order (to avoid effects of magnetostriction). Crystal-field splitting parameters were obtained from non-magnetic DFT calculations (computing barycenters of corresponding bands) using Linearized Muffin-Tin Orbitals method.[17]

RESULTS AND DISCUSSION

**Synthesis, chemical materials properties.** $Na_9Bi_5Os_3O_{24}$ was obtained by reacting the binary constituent oxides at high oxygen pressure and hydrothermal conditions. Black reflective hexagonal blocks or thick plates were harvested after washing the product with water and rinsing with ethanol (Figure 1). Several crystals from different batches were quantitatively analyzed using SEM-EDX, which showed an approximate metal ratio of Na/Bi/Os = 3.1/1.8/1 (Figure S1 in SI). A composition of $Na_{8.63}Bi_{4.81}Os_{3.0}O_{24.4}$ was estimated from ICP-OES and AAS analysis of the metal concentrations (see Table S1 in SI) basically confirming the composition obtained from EDAX. The compound starts decomposing in a stepwise manner at 600 K (See Figure S2 in SI) but the decomposition is not complete till 1273 K.

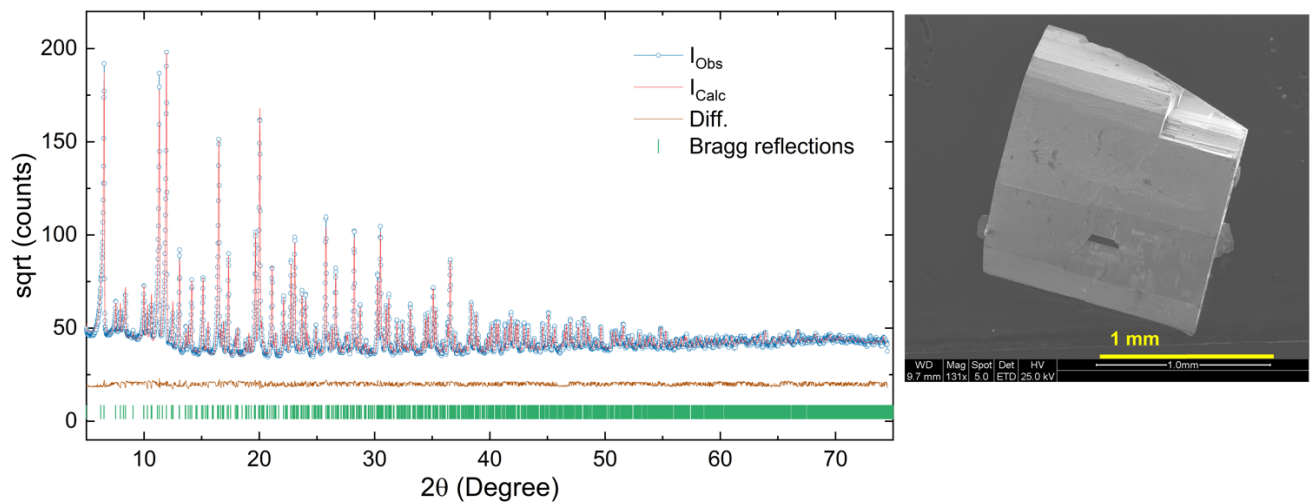

**Figure 1**. Rietveld fit to the powder X-ray diffraction pattern of $Na_9Bi_5Os_3O_{24}$. Blue, Red and brown lines are the observed, calculated and difference curve, respectively. The green vertical bars represent the allowed Bragg reflections. To the right is the electron microscopic image of a typical crystal.

The powder X-ray diffraction pattern was indexed and refined using the final atomic parameters from the single crystal structure determination as a starting model. No apparent impurity phase was observed in the powder pattern. Figure 1 shows the result of a Rietveld profile refinement, for more details see Supporting Information (SI), Tables S2 and S3.

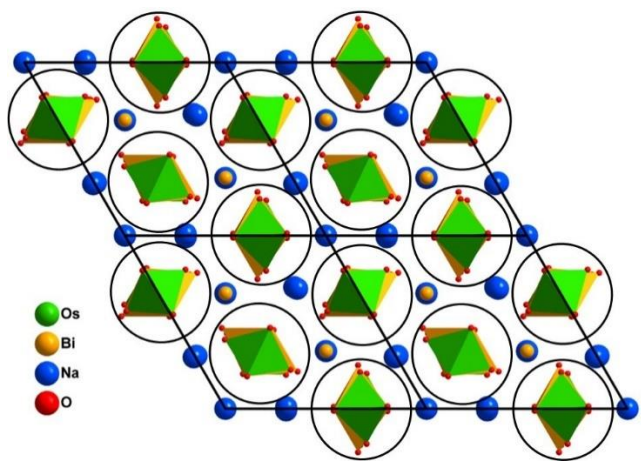

**Figure 2.** Projection of the crystal structure, view direction [001], strands of alternating $OsO_6$ and $BiO_6$ octahedra are emphasized by black circles.

**Crystal structure, description**. From the structure and chemical formula it was deduced that osmium is hexavalent and bismuth is mixed-valent with three out of five (Bi1) being pentavalent and the remaining trivalent (Bi2). $Na_9Bi_5Os_3O_{24}$ exhibits a singular and peculiar crystal structure that cannot be related to any of the prominent aristotypes of typical oxide structure families. Here, for providing a comprehensive description, we resort to the tool of analyzing extended structures in terms of rod packings.[18,19] As is emphasized in Figure 2, all secondary building units are quasi-1D and form chains extending in *c* direction of the hexagonal crystal system. Strands of edge sharing octahedra, centered in an alternating fashion by $Os^{6+}$ and $Bi^{5+}$, are dominating and constitute a precisely hexagonal rod packing. All oxygen atoms in the structure are attached to these chains. Positions and orientations of the latter generate columns of trigonal prisms around the threefold axes of the space group at 1/3 2/3 z (1), 2/3 1/3 z (2) and 0 0 z (3), while a fourth column of oxygen atoms around Na2 is not subject to space group symmetry constraints. The individual rods are shown in Figure 3. The respective local coordinations for the chain of edge sharing octahedra show a typical pattern for $Bi^{5+}$, however, an extremely compressed octahedron around $Os^{6+}$, which is a conspicuous structural feature for this ion and is without precedent. Interestingly, the trigonal prisms of columns (1) and (2) are each occupied by trivalent Bi2 and Na3 in a pairwise alternating fashion, instead of simply alternating, which would be favorable for electrostatic reasons. The coordination polyhedra observed for these atoms are basically as expected. Bismuth(III) forms a trigonal pyramid with three short bonds, and three distinctly longer ones, in accord with the "active lone-pair" scenario.[20,21] Na3 as well shows common distances towards oxygen on average, however, is seemingly unforced in an off-center position. The trigonal strands around the origin of the unit cell comprise alternating filled, $(Na1)O_6$, and empty regular trigonal prisms. Within the remaining chain, Na2 is coordinated by six nearest oxygen atoms, forming a substantially distorted octahedron. These polyhedra are connected to chains oriented along [001] via common edges O1–O2 and O3–O3, respectively. The rods are linked to form a three-dimensional framework, basically via the apical oxygen atoms of the $Bi^{5+}$ and $Os^{6+}$ octahedra. Now, the key role played by the distorted trigonal prisms around trivalent Bi2 and Na3 in controlling the bond length to the apical oxygen atoms is becoming evident: The particular sequence of Bi2 and Na3 is generating a breathing mode of the oxygen atoms reflected by small triangles between Bi2 and Na3 atoms and wider ones between each two subsequent Bi2 and Na3, respectively.

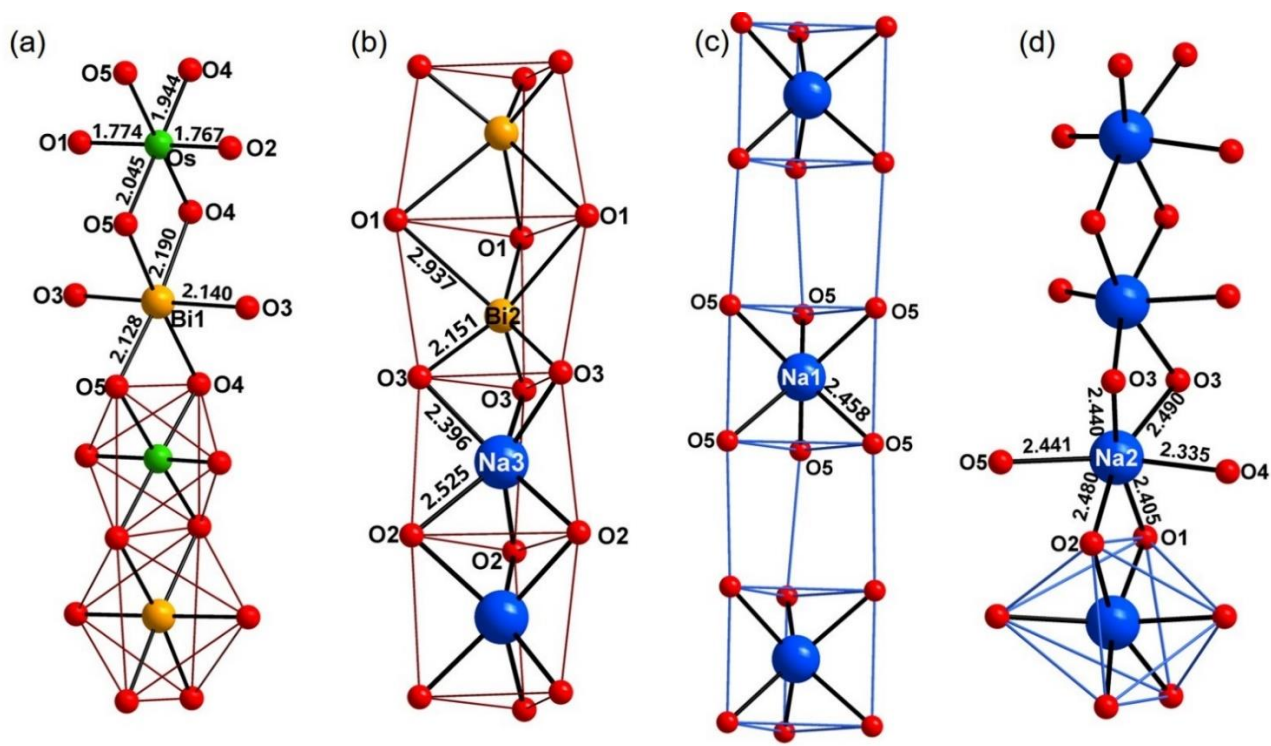

**Figure 3**. Presentation of the individual quasi-1D secondary building units, (a) chains of edge-shared alternating Os/Bi octahedra, (b) chains of alternating pairs of trigonal prisms of Na3/Bi2, (c) chains of alternating empty and filled (Na1)O$_6$ trigonal prisms and (d) chains of edge shared distorted (Na2)O$_6$ octahedra, for details compare text. Bond lengths are in Å.

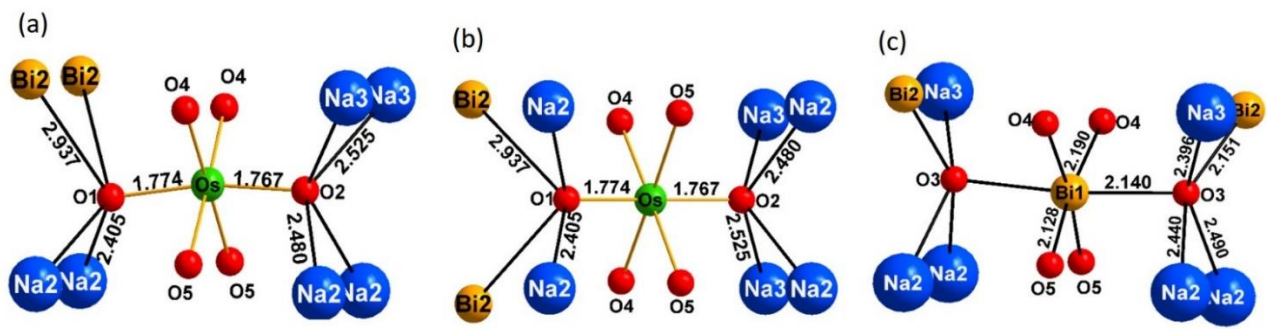

**Figure 4**. First coordination spheres of osmium(VI), (a) view direction perpendicular to, (b) within the crystallographic mirror plane, and (c) of bismuth(V). For osmium, the apical oxygen atoms are in a different local environment, generating local dipole moments, which completely compensate due to space group symmetry. Bond lengths are in Å

As one of the peculiar characteristics of the crystal structure found, cations as different as Bi$^{3+}$ and Na$^+$ (Bi2 and Na3) are hosted in virtually the same kind of trigonal prisms, see Figure 3b. Since this might cause some anti-site disorder between these two positions, we refined the respective site occupation factors. Indeed, slight indications were obtained for a low degree of anti-site disorder, exclusively for these two positions. The site occupation factor (SOF) of 96.53 and 3.47 (%) obtained is at the limits of experimental reliability, since SOF and temperature parameters are commonly strongly correlated in the refinements. This finding does not question the characteristics of the crystal structure, in particular not the short bonds between osmium and the apical oxygen atoms, since the anti-site disorder only inverts the two different surroundings of the apical oxygen atoms. At screening crystals from different batches we became aware that this disorder may vary from one synthesis batch to the other, depending on the temperature schedules applied. At the extreme, SOF of 39.58% to 60.42% was encountered. Data of the split atom refinements are included in SI as Tables S5 – S9. The crystallographic data have been deposited at ICSD under CSD-No. 2063725 and 2063497, respectively.

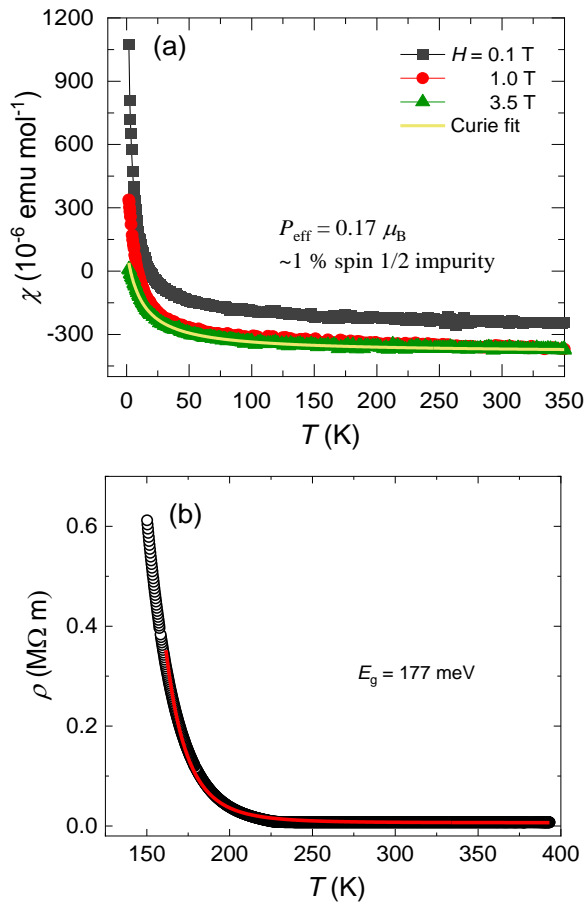

**Figure 5**. Temperature dependent magnetic susceptibility and its Curie fit ($\chi = C/T$) (a) and Arrhenius fit of electrical resistivity (b) of $Na_9Bi_5Os_3O_{24}$ single crystal(s).

The oxygen atoms constituting the smaller triangles are giving space to $Bi^{5+}$ to adopt a virtually regular octahedron, while the wider triangles enable $Os^{6+}$ to develop the strikingly short bonds to the apical oxygen atoms, i.e. to form a substantially compressed octahedron. Interestingly, this contraction is not forced by any matrix effect of the crystal structure. From Figure 4 one can see that there is sufficient freedom within the linkages Os-O1-Bi2 and Os-O2-Na3 to shift the respective oxygen atoms away from osmium. Even in the extreme, when moving O2 into the barycenter of the quadrilateral formed by Na2(2×) and Na3(2×), the Na-O separations are still within the common limit, see Figure S3. Thus, from a crystal chemistry point of view, there must be local electronic effects that are driving osmium into that peculiar configuration found experimentally.

**Physical properties**. Figure 5(a) shows the magnetic susceptibility carried out in the temperature range of 2-350 K at different applied magnetic fields (0.1, 1.0, 3.5 T), clearly indicating the diamagnetic nature of the sample. Minor Curie tailing observed at low temperatures corresponds to an upper estimate of a magnetic moment $P_{eff} = 0.17$ $\mu_B$/mol corresponding to 1 % of a spin ½ impurity (from a Curie fit of the susceptibility in the entire temperature range). Electrical resistivity data (Figure 5b) shows that the compound is semiconducting with a band gap of 177 meV.

**Computational results, theoretical analyses**. In order to uncover the physical origin of the low-spin state stabilization in $Na_9Bi_5Os_3O_{24}$ we calculated the crystal-field splitting of the Os $t_{2g}$ sub-shell in the DFT approach (see Figure 6) and found that it amounts to 1.4 eV, with the singlet $xy$ orbital lying below the doublet $xz$, $yz$ (in the local coordinate system with axes directed from Os towards oxygen atoms, $z$ direction to the apical ligands). This separation virtually corresponds by magnitude to the $t_{2g}$-$e_g$ splitting (*10Dq*) in conventional *3d* transition metal oxides. Considering Hund's energy for *5d* ions of ~ 0.5 eV,[22] it is well understandable that the lowest $xy$ orbital is filled by two electrons of $Os^{6+}$, resulting in a singlet ground state, *S=0*. Thus, it is clearly the strong crystal-field (ratio of short and long Os-O bonds is ~0.89), which leads to the non-magnetic state in the investigated material. It is worthwhile mentioning that, based on various experimental evidences, there have been claims of a low-spin state realized in $t_{2g}$ configurations.[6] However, respective theoretical and experimental studies did not confirm such a picture, see[7,22] and references therein. To the best of our knowledge $Na_9Bi_5Os_3O_{24}$ is thus the only oxide, where such a situation affecting a set of $t_{2g}$ levels is actually realized.

For checking the role of the degree of filling of the *5d* derived states, we performed crystal structure optimizations with an artificially reduced number of electrons in the system of six per formula unit (compensated by a corresponding surrounding charge) or by substituting $W^{6+}$ ($5d^0$) for $Os^{6+}$ (volume, unit cell shape, atomic positions were allowed to relax). The results are qualitatively similar and rather illustrative. In the first case, of reduced number of electrons, the distance from the plane formed by equatorial oxygen atoms to apical ligands increases (1.87 and 1.97 Å), while the in-plane M-O bond lengths decrease (1.87 Å(2×) and 1.89 Å(2×)), for positional parameters see Table S10 in SI. When giving the $5d^2$ electrons back to the system (or reinstating Os for W), the structure returns to the original experimental configuration upon computational relaxation. These observations clearly demonstrate that the compression of the $OsO_6$ octahedron is basically caused by electronic effects, *i.e.* is due to a JT effect lifting the $t_{2g}$ orbital degeneracy.

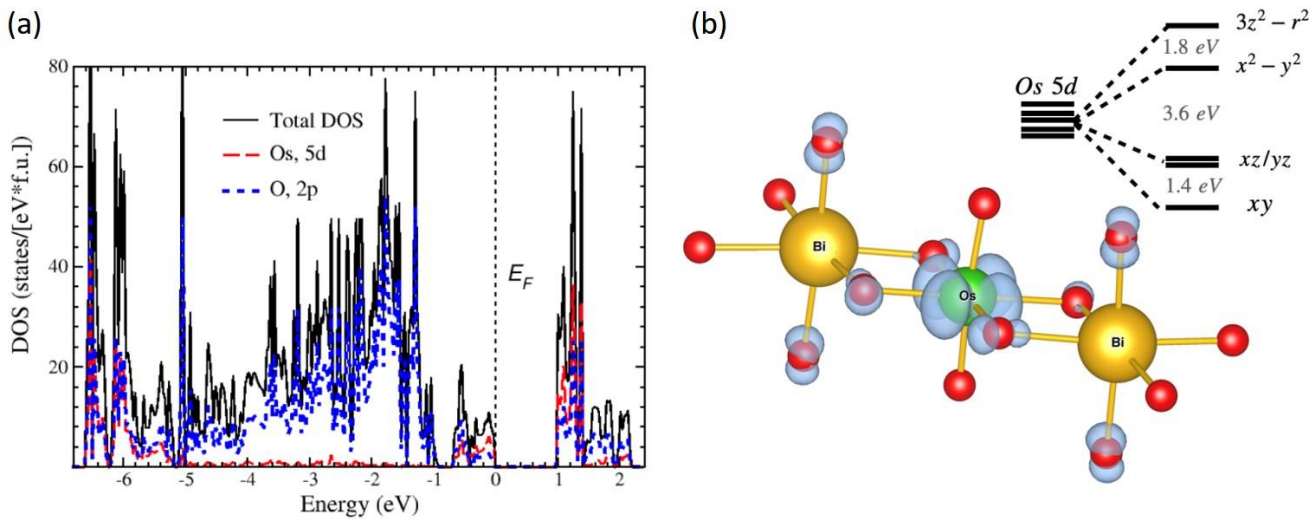

**Figure 6.** Total and partial density of states plots (a) and charge density (blue) corresponding to a single occupied band (just below the fermi level, see left panel) in the GGA+U calculation (b). This band has predominantly Os $d$ states character. Osmium (oxygen) atoms are shown by green (red) balls. Inset of (b) shows Os $5d$ level scheme.

It appears advisable to compare the situation in $Na_9Bi_5Os_3O_{24}$ with other functional oxides containing $Os^{6+}$. In principle, open shell $5d$ systems can arrive at a nonmagnetic state due to strong spin-orbit interaction even for a cubic crystal field. This has been encountered for $5d^4$ ions like $Ir^{5+}$.[23,24] Also for $Os^{6+}$ ($5d^2$) a nonmagnetic ground state owing to a combined action of spin-orbit coupling and cubic crystal field (10 $Dq$) has been reported.[25-27] The character of this state, however, is very different from what we have in $Na_9Bi_5Os_3O_{24}$: it is not a singlet state with $S=0$, but a non-Kramers $e_g$ doublet. Such an electronic state gives rise to low-lying magnetic states (for double perovskites like $Ba_2MOsO_6$, M=Ca, Mg, Zn the singlet-triplet gap is estimated as ~25 meV).[25] This would give strong and temperature-dependent van Vleck paramagnetism. In our case the situation is fundamentally different: there is a singlet state of $Os^{6+}$ with $S=0$, with the gap to the lowest-lying magnetic state ($\Delta_{CF} - J_{Hund}$) = ~ 1 eV. Spin-orbit coupling plays practically no role in its stabilization. DFT+U+SOC calculations fully support this conclusion. The orbital moment was found to be extremely small, ~$10^{-3}$ $\mu_B$.

CONCLUSIONS

We report on new $Na_9Bi_5Os_3O_{24}$ which features a unique crystal structure and an unparalleled electronic ground state: an extremely compressed octahedron of the oxygen atoms coordinating osmium(VI) is giving rise to a nonmagnetic configuration by hosting the $5d^2$ electrons paired in the $d_{xy}$ derived state. The splitting of the $t_{2g}$ levels amounts to 1.4 eV, substantially surpassing the Hund's coupling energy for $5d$ ions of ~ 0.5 eV. This kind of a giant JT distortion, which is without precedent, is not enforced by structural matrix effects: the structure would offer sufficient space to enable a virtually regular octahedron without causing strain. Apparently, the energy landscape of configurations for $5d^2$ systems, like $Os^{6+}$, in approximately octahedral coordination by oxygen is rather flat and, depending on the balance of local electronic factors of influence vs. structural constraints, diverse ground states may develop. Against this background it appears rewarding to inspect this kind of competing impacts, structural frustration vs. strive for an as low as possible local electronic state, on the basis of the structure type presented, which offers plenty of options for modifying the relevant conditions by aliovalent or isovalent substitution of any of the cations present. Such computational studies, at best followed by experimental validation, would have the potential to provide valuable insights in the hierarchy of perturbations defining the ground states of open shell transition element compounds.

## ASSOCIATED CONTENT

**Supporting Information**. Results of quantitative analysis (EDS and wet chemical), thermal decomposition profile, additional structure figures, tables for crystallographic data, atomic coordinates, bond distances and anisotropic displacement parameters. "This material is available free of charge via the Internet at http://pubs.acs.org."

## AUTHOR INFORMATION

**Corresponding Author**

\* Prof. Dr. Martin Jansen
*Email: m.jansen@fkf.mpg.de*

**Author Contributions**

All authors have given approval to the final version of the manuscript.

## ACKNOWLEDGMENT

G.S.T. thank the Cluster of Excellence *ct.qmat* (EXC 2147) funded by the Deutsche Forschungsgemeinschaft (DFG) for partial support. A.V.U. is grateful to the Quantum project (AAAA-A18-118020190095-4). The work of D. Kh. was funded by the DFG (German Research Foundation) - Project number 277146847 - CRC 1238. DFT+U calculations (S.V.S.) were supported by the Russian Science Foundation via RSF-20-62-46047 project.

*Supplementary Material*

# Na9Bi5Os3O24: A Unique Diamagnetic Oxide Featuring a Pronouncedly Jahn-Teller Compressed Octahedral Coordination of Osmium(VI)

Gohil S. Thakur,[1,2] Hans Reuter,[3] Alexey V. Ushakov,[4] Gianpiero Gallo,[5] Jürgen Nuss,[5] Robert E. Dinnebier,[5] Sergey V. Streltsov,[4] Daniel I. Khomskii,[6] and Martin Jansen[5]*

Contents:

1. Quantitative analysis using SEM-EDX

2. Thermal decomposition profile

3. Additional structures sectionas from experiment and DFT calculations

4. Results of chemical analysis

5. Tables of atomic coordinates, bond lengths and thermal parameter

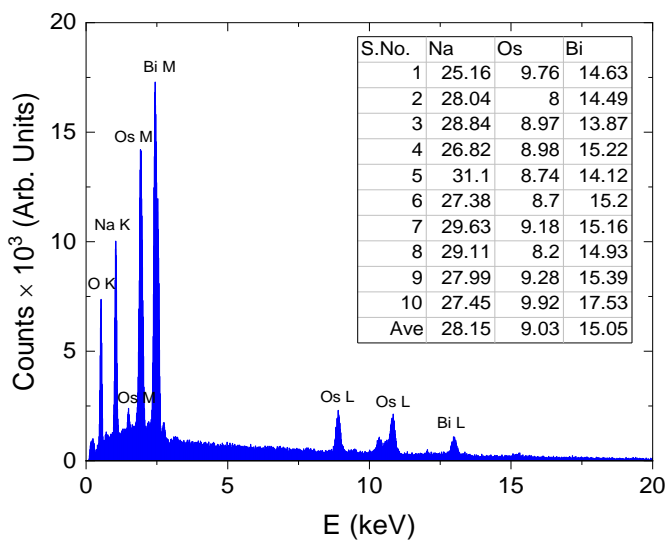

Figure S1: Elemental analysis of a typical crystal of Na$_9$Bi$_5$Os$_3$O$_{24}$ using SEM-EDX. Inset shows table of atomic ratios at different regions.

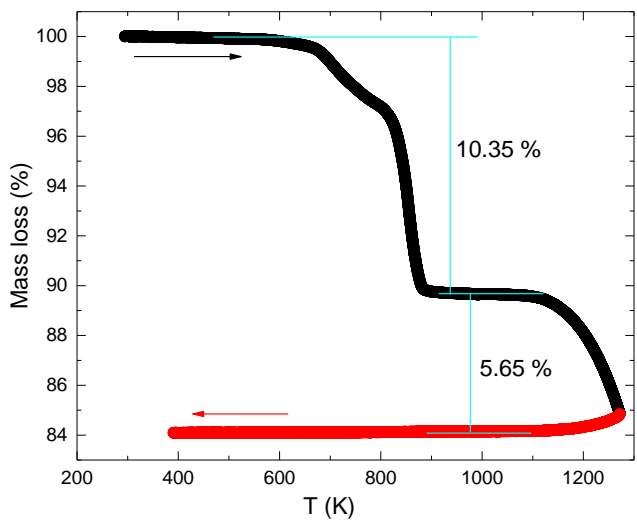

Figure S2. Thermal decomposition profile of Na$_9$Bi$_5$Os$_3$O$_{24}$.

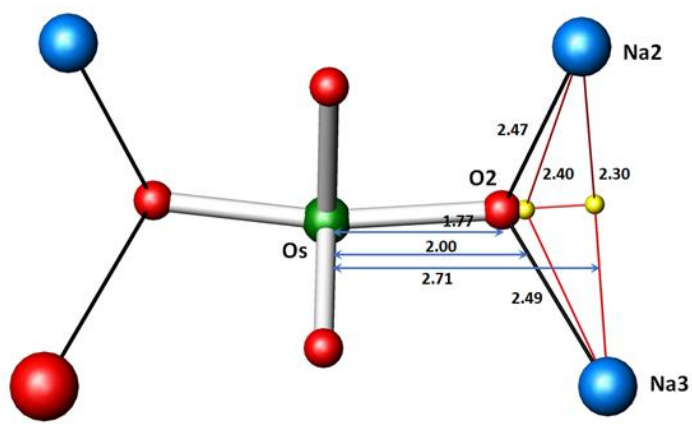

Figure S3. Demonstrating lack of structural frustration with respect to short apical Os-O2 bond. The distances are in Å.

Table S1. Wet chemical (ICP-OES/ AAS) analysis. Mole fraction is normalized to Os.

| Atom | Mass % (1 run) | Mass % (2 run) | Average | Mole fraction | Normalized |
|---|---|---|---|---|---|
| Bi | 46.5 | 46.4 | 46.45 | 0.2223 | 4.8 |
| Na | 9.34 | 8.98 | 9.16 | 0.3984 | 8.6 |
| Os | 26.5 | 26.2 | 26.35 | 0.1385 | 3.0 |
| O (remaining) | 17.66 | 18.42 | 18.04 | 1.128 | 24.4 |

Table S2: Atomic coordinates for Na$_9$Bi$_5$Os$_3$O$_{24}$ obtained by Rietveld refinement of room temperature powder data.

| Atom | Wyck. | x | y | z |
|---|---|---|---|---|
| Os1 | 6h | 0.66806 | -0.0022 | 1/4 |
| Bi1 | 6g | 0.67116 | 0 | 1/2 |
| Bi2 | 4f | 1/3 | 2/3 | 0.606 |
| Na1 | 2b | 0 | 0 | 1/4 |
| Na2 | 12i | 0.29206 | 0.98487 | 0.37917 |
| Na3 | 4f | 1/3 | 2/3 | 0.39554 |
| O1 | 6h | 0.80232 | 0.1869 | 1/4 |
| O2 | 6h | 0.56344 | 0.79511 | 1/4 |
| O3 | 12i | 0.52977 | 0.75106 | 0.49004 |
| O4 | 12i | 0.52719 | 0.00702 | 0.37432 |
| O5 | 12i | 0.78861 | 0.99129 | 0.38295 |

Table S3: Selected bond distances obtained for $Na_9Bi_5Os_3O_{24}$ obtained by Rietveld refinement.

| Atom pairs | *Distance* (Å) | Atom pairs | *Distance* (Å) |
|---|---|---|---|
| Os1-O1 (×1) | 1.656 | Na1-O5 (×6) | 2.658 |
| Os1-O2 (×1) | 1.725 | | |
| Os1-O5 (×2) | 2.099 | Na2-O4 (×1) | 2.211 |
| Os1-O4 (×2) | 2.146 | Na2-O3 (×1) | 2.337 |
| | | Na2-O5 (×1) | 2.453 |
| Bi1-O5 (×2) | 1.925 | Na2-O1 (×1) | 2.461 |
| Bi1-O3 (×2) | 2.129 | Na2-O2 (×1) | 2.490 |
| Bi1-O4 (×2) | 2.171 | Na2-O3 (×1) | 2.688 |
| | | | |
| Bi2-O3 (×3) | 2.243 | Na3-O3 (×3) | 2.071 |
| Bi2-O4 (×3) | 2.798 | Na3-O2 (×3) | 2.711 |

Table S4: Anisotropic displacement parameters (Å$^2$ x 10$^3$) for $Na_9Bi_5Os_3O_{24}$, ordered structure model. The anisotropic displacement factor exponent takes the form: $-2\pi^2[h^2a^{*2}U_{11} + ... + 2hka^*b^*U_{12}]$.

| Atom | U11 | U22 | U33 | U23 | U13 | U12 |
|---|---|---|---|---|---|---|
| Os | 5(1) | 5(1) | 2(1) | 0 | 0 | 2(1) |
| Bi1 | 6(1) | 5(1) | 2(1) | 0(1) | 0(1) | 2(1) |
| Bi2 | 9(1) | 9(1) | 5(1) | 0 | 0 | 4(1) |
| Na1 | 15(2) | 15(2) | 20(3) | 0 | 0 | 7(1) |
| Na2 | 9(1) | 14(2) | 12(2) | 1(1) | 0(1) | 3(1) |
| Na3 | 3(1) | 3(1) | 43(4) | 0 | 0 | 1(1) |
| O1 | 14(4) | 3(3) | 9(2) | 0 | 0 | 2(2) |
| O2 | 15(3) | 2(3) | 9(2) | 0 | 0 | 1(3) |
| O3 | 9(2) | 5(2) | 8(2) | 1(2) | 2(2) | 4(2) |
| O4 | 6(2) | 14(2) | 6(2) | -1(2) | 1(2) | 6(2) |
| O5 | 9(2) | 14(2) | 4(2) | 1(2) | 0(2) | 7(2) |

Table S5: Crystal data and structure refinement for $Na_9Bi^V_3Bi^{III}_2Os_3O_{24}$.

| Crystal | 1 | 2 |
|---|---|---|
| | | |
| Identification code | HR2212_0m | nus313a_0m |
| Empirical formula | $Bi_5Na_9O_{24}Os_3$ | |
| Formula weight [g/mol] | 2206.41 | |
| T [K] | 296(2) | |
| *Crystal data* | | |
| **Crystal system, space group** | Hexagonal, $P\bar{6}2c$ | |
| **a [Å]** | 9.8115(3) | 9.8344(2) |
| **c [Å]** | 12.8457(5) | 12.8732(5) |
| **V [Å$^3$]** | 1070.93(8) | 1078.23(6) |
| **Z, d$_{calc}$ [g/cm$^3$]** | 2, 6.842 | 2, 6.796 |
| **μ(MoK$_a$) [mm$^{-1}$]** | 58,944 | 58,545 |
| **F(000)** | 1868 | |
| *Data collection* | | |
| **2Θ$_{max}$** | 70° | |
| **Reflections collected** | 75234 | 18635 |
| **Reflections unique, R$_{int}$** | 1637, 0.0877 | 1639, 0.0508 |
| *Refinement* | | | |
| **Data / restraints / parameters** | 1637 / 0 / 70 | 1637 / 0 / 75 | 1639 / 0 / 76 |
| **Goodness-of-fit on F$^2$** | 1,040 | 1,037 | 1,055 |
| **R$_1$, wR2 [I > 2s(I)]** | 0.0208, 0.0529 | 0.0176, 0.0421 | 0.0242, 0.0911 |
| **R$_1$, wR2 [all data]** | 0.0268, 0.0567 | 0.0234, 0.0451 | 0.0288, 0.0965 |
| **Absolute structure parameter** | -0.038(6) | -0.033(11) | 0.024(11) |
| **Extinction coefficient** | 0.00102(5) | 0.00144(6) | 0.00076(10) |
| **±D [eÅ$^{-3}$]** | 1.729/-3.396 | 1.728/-1.905 | 2.687/-3.289 |
| *Structure model* | | | |
| **Na/Bi-disorder** | no | yes | yes |
| **Degree of anti-site disorder** | 0% | 96.53%/3.47% | 39.58%/60.42% |

Table S6. Atomic coordinates (x 10$^4$) and equivalent isotropic displacement parameters (Å$^2$ x 10$^3$) for crystal 1 and 2 of Na$_9$Bi$_5$Os$_3$O$_{24}$ obtained after refinement (top rows) and with anti-site disorder (bottom rows). U(eq) is defined as one third of the trace of the orthogonalized U$_{ij}$ tensor.

| Atom | Sample | Bi/Na-disorder | Wyckoff | Occ. | x | y | z | U(eq) |
|---|---|---|---|---|---|---|---|---|
| Os | 1 | no | 6h | | 6682(1) | 9975(1) | 2500 | 4(1) |
| | 1 | yes | | | 6682(1) | 9975(1) | 2500 | 5(1) |
| | 2 | yes | | | 3300(1) | -8(1) | 2500 | 9(1) |
| Bi1 | 1 | no | 6g | | 6712(1) | 10000 | 5000 | 5(1) |
| | 1 | yes | | | 6711(1) | 10000 | 5000 | 5(1) |
| | 2 | yes | | | 3294(1) | 0 | 5000 | 10(1) |
| Bi2A | 1 | no | 4f | | 3333 | 6667 | 6051(1) | 7(1) |
| | 1 | yes | | 0.965(3) | 3333 | 6667 | 6051(1) | 7(1) |
| | 2 | yes | | 0.396(3) | 6667 | 3333 | 3961(1) | 16(1) |
| Bi2B | 1 | no | | | | | | |

| Atom | Crystal | Na/Bi-disorder | Wyckoff | Occupancy | x | y | z | Ueq |
|---|---|---|---|---|---|---|---|---|
| | 1 | yes | 4f | 0.035 | 3333 | 6667 | 3997(11) | 11(3) |
| | 2 | yes | | 0.604 | 6667 | 3333 | 6043(1) | 10(1) |
| Na1 | 1 | no | 2b | | 10000 | 10000 | 2500 | 17(1) |
| | 1 | yes | | | 10000 | 10000 | 2500 | 19(1) |
| | 2 | yes | | | 0 | 0 | 7500 | 28(2) |
| Na2 | 1 | no | 12i | | 2976(6) | 9912(4) | 3789(2) | 13(1) |
| | 1 | yes | | | 2994(5) | 9916(3) | 3788(2) | 13(1) |
| | 2 | yes | | | 6949(8) | -24(4) | 6211(3) | 20(1) |
| Na3A | 1 | no | 4f | | 3333 | 6667 | 3731(4) | 16(1) |
| | 1 | yes | | 0.965(3) | 3333 | 6667 | 3654(6) | 19(2) |
| | 2 | yes | | 0.396(3) | 6667 | 3333 | 6370(20) | 57(13) |
| Na3B | 1 | | | | | | | |
| | 1 | yes | 4f | 0.035 | 3333 | 6667 | 6440(140) | 19(2) |
| | 2 | yes | | 0.604 | 6667 | 3333 | 3581(15) | 47(6) |
| O1 | 1 | no | 6h | | 8035(13) | 12028(12) | 2500 | 10(1) |
| | 1 | yes | | | 8032(12) | 12026(11) | 2500 | 12(1) |
| | 2 | yes | | | 2159(18) | -2079(15) | 7500 | 22(2) |
| O2 | 1 | no | 6h | | 5648(8) | 7895(11) | 2500 | 10(1) |
| | 1 | yes | | | 5649(7) | 7892(10) | 2500 | 12(1) |
| | 2 | yes | | | 4127(11) | 2040(14) | 7500 | 22(2) |
| O3 | 1 | no | 12i | | 5336(5) | 7486(6) | 5037(3) | 7(1) |
| | 1 | yes | | | 5337(5) | 7490(5) | 5037(3) | 9(1) |
| | 2 | yes | | | 4649(6) | 2510(6) | 5010(4) | 11(1) |
| O4 | 1 | no | 12i | | 5443(6) | 10117(6) | 3631(4) | 8(1) |
| | 1 | yes | | | 5443(5) | 10119(5) | 3632(3) | 10(1) |
| | 2 | yes | | | 4624(7) | 32(8) | 6365(5) | 18(1) |
| O5 | 1 | no | 12i | | 7987(7) | 9862(7) | 3703(4) | 8(1) |
| | 1 | yes | | | 7983(6) | 9859(6) | 3703(3) | 10(1) |
| | 2 | yes | | | 1925(8) | -36(8) | 6290(4) | 17(1) |

Table S7. Summary of bond lengths [Å] obtained from single crystal structure refinement using different protocols of refinement for all crystal samples

| Crystal | 1 | 1 | 2 |
|---|---|---|---|
| Na/Bi-disorder | no | yes | yes |
| Degree of disorder | 0% | 96.53%/3.47% | 39.58%/60.42 |
| Os1-O2 (×1) | 1.767(10) | 1.770(9) | 1.755(13) |
| Os1-O1 (×1) | 1.773(11) | 1.771(10) | 1.767(14) |
| Os1-O4 (×2) | 1.944(5) | 1.945(4) | 1.944(6) |
| Os1-O5 (×2) | 2.044(5) | 2.043(4) | 2.054(6) |
| Bi1-O5 (×2) | 2.129(5) | 2.128(4) | 2.126(6) |
| Bi1-O3 (×2) | 2.140(5) | 2.136(4) | 2.140(5) |
| Bi1-O4 (×2) | 2.191(5) | 2.190(4) | 2.182(6) |
| Bi2A-O3 (×3) | 2.150(5) | 2.151(4) | 2.193(5) |
| Bi2B-O3 (×3) | | 2.172(10) | 2.180(5) |
| Na1-O5 (×6) | 2.457(5) | 2.459(4) | 2.466(7) |
| Na2-O4 (×1) | 2.335(7) | 2.319(6) | 2.323(10) |
| Na2-O1 (×1) | 2.406(9) | 2.404(8) | 2.465(12) |
| Na2-O3 (×1) | 2.440(6) | 2.440(5) | 2.463(7) |
| Na2-O5 (×1) | 2.441(7) | 2.453(6) | 2.580(9) |
| Na2-O2 (×1) | 2.480(8) | 2.478(7) | 2.447(12) |
| Na2-O3 (×1) | 2.490(6) | 2.487(5) | 2.450(6) |
| Na2-O5 (×1) | | | 2.667(9) |
| Na3A-O2 (×3) | 2.396(6) | 2.464(6) | 2.608(18) |
| Na3A-O3 (×3) | 2.525(6) | 2.467(7) | 2.46(2) |
| Na3B-O3 (×3) | | 2.49(13) | 2.524(15) |
| Na3B-O1 (×3) | | 2.65(9) | 2.493(13) |
| | | | |
| | = Affected by disorder | | |

Table S8. Anisotropic displacement parameters (Å$^2$ × 10$^3$) for Na$_9$Bi$_5$Os$_3$O$_{24}$ assuming anti-site disorder (crystal 1). The anisotropic displacement factor exponent takes the form: $-2\pi^2[h^2a^{*2} U_{11} + ... + 2hka^*b^*U_{12}]$

| Atom | U11 | U22 | U33 | U23 | U13 | U12 |
|---|---|---|---|---|---|---|
| Os1 | 6(1) | 5(1) | 3(1) | 0 | 0 | 3(1) |
| Bi1 | 7(1) | 6(1) | 3(1) | 0(1) | 0(1) | 3(1) |
| Bi2A | 9(1) | 9(1) | 5(1) | 0 | 0 | 4(1) |
| Bi2B | 12(3) | 12(3) | 8(7) | 0 | 0 | 6(2) |
| Na1 | 17(2) | 17(2) | 23(3) | 0 | 0 | 9(1) |
| Na2 | 12(1) | 12(2) | 12(2) | 0(1) | -1(1) | 4(1) |
| Na3A | 19(2) | 19(2) | 19(3) | 0 | 0 | 10(1) |
| Na3B | 19(2) | 19(2) | 19(3) | 0 | 0 | 10(1) |
| O1 | 15(4) | 6(3) | 10(2) | 0 | 0 | 3(2) |
| O2 | 17(3) | 5(3) | 10(2) | 0 | 0 | 3(3) |
| O3 | 10(2) | 7(2) | 10(2) | 1(2) | 3(1) | 4(2) |
| 4O4 | 8(2) | 15(2) | 8(2) | -1(2) | 0(1) | 7(2) |
| O5 | 11(2) | 16(2) | 5(2) | 1(2) | 1(1) | 8(2) |

Table S9. Anisotropic displacement parameters (Å$^2$ × 10$^3$) for Na$_9$Bi$_5$Os$_3$O$_{24}$ assuming inversion and disorder (crystal 2). The anisotropic displacement factor exponent takes the form: $-2\pi^2[h^2a^{*2} U_{11} + ... + 2hka^*b^*U_{12}]$

| Atom | U11 | U22 | U33 | U23 | U13 | U12 |
|---|---|---|---|---|---|---|
| Os1 | 10(1) | 10(1) | 8(1) | 0 | 0 | 5(1) |
| Bi1 | 12(1) | 10(1) | 8(1) | 0(1) | 0(1) | 5(1) |
| Bi2A | 16(1) | 16(1) | 16(1) | 0 | 0 | 8(1) |
| Bi2B | 12(1) | 12(1) | 9 (1) | 0 | 0 | 6(1) |
| Na1 | 21(2) | 21(2) | 43(5) | 0 | 0 | 10(1) |
| Na2 | 21(2) | 16(2) | 22(2) | 1(1) | 0(1) | 8(2) |
| Na3A | 80(19) | 80(19) | 13(10) | 0 | 0 | 40(10) |
| Na3B | 62(9) | 62(9) | 15(6) | 0 | 0 | 31(4) |
| O1 | 30(6) | 11(4) | 15(3) | 0 | 0 | 3(3) |
| O2 | 37(4) | 5(4) | 16(3) | 0 | 0 | 5(4) |
| O3 | 15(2) | 5(2) | 14(2) | 0(2) | 0(1) | 5(2) |
| O4 | 15(3) | 26(3) | 15(2) | -1(2) | 0(2) | 13(2) |
| O5 | 18(3) | 27(3) | 9(2) | 1(2) | 1(2) | 14(3) |

Table S10. DFT+U optimized crystal structure (unit cell volume and shape, atomic positions were allowed to relax, number of electrons decreased by 2 per each Os (compensated by external charges). Space group: $P\bar{6}2c$, $a$ = 9.01577 Å, $c$ = 12.38993 Å.

| Atom | Wyckoff | x | y | z |
|---|---|---|---|---|
| Os1 | 6h | 0.66543 | -0.02727 | 0.2500 |
| Bi1 | 6g | 0.66530) | 0.0000 | 0.000 |
| Bi2 | 4f | 0.3333 | 0.6667 | 0.62392 |
| Na1 | 2b | 0.0000 | 0.0000 | 0.2500 |
| Na2 | 12i | 0.28790 | -0.0023 | 0.37657 |
| Na3 | 4f | 0.3333 | 0.6667 | 0.37511 |
| O1 | 6h | 0.77653 | 0.24262 | 0.2500 |
| O2 | 61h | 0.54505 | 0.75121 | 0.2500 |
| O3 | 12i | 0.53331 | 0.73165 | 0.50826 |
| O4 | 12i | 0.52839 | 0.00057 | 0.35882 |
| O5 | 12l | 0.81560 | 0.00601 | 0.35957 |